# Exact solutions for spherically gravitational collapse around a black hole: the effect of tangential pressure[*]

Zhao Sheng-Xian(赵声贤)[1,3], Zhang Shuang-Nan(张双南)[1,2,3,†]

[1] *Key Laboratory of Space Astronomy and Technology, National Astronomical Observatories, Chinese Academy of Sciences, Beijing 100012, China*

[2] *Key Laboratory of Particle Astrophysics, Institute of High Energy Physics, Beijing 100049, China*

[3] *University of Chinese Academy of Sciences, Beijing 100049, China*

**Abstract:** Spherically gravitational collapse towards a black hole with non-zero tangential pressure is studied. Exact solutions corresponding to different equations of state are given. We find that when taking the tangential pressure into account, the exact solutions have three qualitatively different endings. For positive tangential pressure, the shell around a black hole may eventually collapse onto the black hole, or expand to infinity, or have a static but unstable solution, depending on the combination of black hole mass, mass of the shell and the pressure parameter. For vanishing or negative pressure, the shell will collapse onto the black hole. For all eventually collapsing solutions, the shell will cross the event horizon, instead of accumulating outside the event horizon, even if clocked by a distant stationary observer.

## 1. Introduction

In 1939 Oppenheimer and Snyder [1] studied the gravitational collapse of a homogeneous spherically symmetric and pressureless dust cloud, which initiated the study of gravitational collapse. In the work of Oppenheimer and Snyder (1939) [1], they predicted the phenomenon of "frozen star", which states that a test particle falling towards a black hole will be eventually frozen near the black hole with an arbitrarily small distance from the event horizon for an observer at infinity, though the particle will indeed cross the event horizon and reach the singularity at the center within finite time in the comoving coordinates. For a long time it has been believed that their work gave a faithful description of gravitational collapse. On the other hand, Liu and Zhang (2009) [2] have found that in the Schwarzschild coordinates the shell around a black hole will actually cross the event horizon of the system when taking account of the mass of the shell.

This problem is closely related to the problem of gravitational collapse in general relativity. Following the original work of Oppenheimer and Snyder, many works have been done on the formation of black holes, e.g. [3-12]. There are also some works concerning collapsing shells [13-19]. Analytical results on the collapse of fluids with pressure are rare, and little is known about the end state of collapse in these systems. For example, Christodoulou [8] introduced a two-phase fluid model with pressure to capture some of the features of actual stellar gravitational collapse, but the problems of the formation of black holes in this situation has not been investigated. Numerical methods also have been used to integrate Einstein's equations. Ori and Piran [20] numerically investigated the gravitational collapse of a perfect fluid with a barotropic equation of state $p = k\rho$, where $p$ and $\rho$ are the pressure and the proper energy density of the fluid respectively, and showed that there are solutions with a naked singularity, as well as black hole solutions for every value of $k$ lies in the range $0 \le k \le 0.4$. In Ori and Piran's work, only self-similar solutions are considered and there is no black hole in the initial data.

[*] Project supported by the National Natural Science Foundation of China under grants 11373036 and 11133002, the National Program on Key Research and Development Project (Grant No. 2016YFA0400802), the Key Research Program of Frontier Sciences, CAS, Grant NO. QYZDY-SSW-SLH008

[†] E-mail: zhangsn@ihep.ac.cn

In Liu and Zhang' work [2], they have studied an ideal model of a spherically symmetric and constant density shell (with finite thickness) collapsing towards a pre-existing black hole locating at the center of spherical symmetry, but their model of matter shell is free of pressure. However, in many real astrophysical settings, the matter shell collapsing towards a black hole has non-vanishing pressure. In this paper, we will deal with more general cases, and study the problem of one fluid shell with finite thickness and non-vanishing pressure collapsing towards a black hole in both comoving coordinates and Schwarzschild coordinates, and compare our results with Liu and Zhang's [2]. To make the problem tractable, we only consider a model of fluid with only tangential pressure, which may serve as a guide to future investigations of more realistic models. The analysis of the Einstein's equations is considerably simplified in this case, and this model has attracted some studies in the past, especially on the issue of naked singularity [11][21-23]. For example, Barve, Singh and Witten [11] showed that if a singularity forms at the center in the tangential pressure model, the conditions for the singularity to be naked are exactly the same as in the model of dust collapse, by using a series solution for the evolution of the area radius in comoving coordinates. Our concerns are the possible fates of a matter shell with tangential pressure around a pre-existing black hole in both comoving coordinates and Schwarzschild coordinates. We find that the exact solutions have three qualitatively different endings. For positive tangential pressure, the shell around a black hole may eventually collapse onto the black hole, or expand to infinity, or have a static but unstable solution. For vanishing or negative pressure, the shell will collapse onto the black hole. For all eventually collapsing solutions, the results are qualitatively similar to Liu and Zhang's, that is the in-falling matter shell of finite thickness with tangential pressure can indeed cross the event horizon of the pre-existing black hole, even if clocked by a distant stationary observer; thus the paradox of frozen star actually is completely resolved even when the tangential pressure is taken into account. We adopt $G = c = 1$ throughout this paper.

## 2. The general solution

We consider a spherically symmetric spacetime with a black hole surrounded by a fluid shell initially outside the apparent horizon and having only tangential pressure.

In the comoving coordinates, the spherically symmetric metric is [24]

$$ds^2 = e^\sigma dt^2 - e^\omega dr^2 - R^2 d\Omega^2, \tag{1}$$

where $\sigma$, $\omega$ and $R$ are functions of $t$ and $r$. In the comoving coordinates, the energy-momentum tensor is [25]

$$T_b^a = (\rho, -p_1, -p_2, -p_3), \tag{2}$$

here $\rho$ is the proper energy density, $p_\alpha (\alpha = 1,2,3)$ represent the principle pressures.

In the case we are considering, $p_1 = 0$ and $p_2 = p_3 = p_T$, that is the radial pressure vanishes. Let $r = a$ and $r = b$ be the inner radius and outer radius of the fluid shell, respectively, the Einstein's equations for this system are [9] [11]

$$\dot{m} = 0, \tag{3}$$

$$m' = 4\pi \rho R^2 R', \tag{4}$$

$$\dot{\omega} = -2\frac{\dot{\rho}}{\rho} - 4\frac{\dot{R}}{R}(1 + \frac{p_T}{\rho}), \tag{5}$$

$$\sigma' = 4\frac{R'}{R}\frac{p_T}{\rho}, \tag{6}$$

and

$$m(t,r) = \frac{1}{2}R(1 + e^{-\sigma}\dot{R}^2 - e^{-\omega}R'^2), \tag{7}$$

here we use dots and primes to indicate the partial derivatives with respect to $t$ and $r$, respectively. Here $m$ can be interpreted as an energy, which includes contributions from the kinetic energy and the gravitational potential energy. From equation (3) we have

$$m = m(r). \tag{8}$$

Take

$$p_T = k\rho \tag{9}$$

as the equation of state of the shell. The weak energy condition will hold if $\rho \geq 0$ and $-1 \leq k \leq 1$. Integrate equations (4) and (5) by virtue of this, we obtain

$$e^{-\omega(t,r)} = C(r)\frac{R^{4k}}{R'^2}. \tag{10}$$

Substitute this equation into (7), we then have

$$\left(\frac{dR}{d\tau}\right)^2 = \frac{2m}{R} - 1 + C(r)R^{4k}, \tag{11}$$

where $d\tau = e^{\sigma/2}dt$ is the proper time.

Rewrite equation (11) as

$$(\frac{dR}{d\tau})^2 + V(R) = -1, \tag{12}$$

where

$$V(R) = -(\frac{2m}{R} + C(r)R^{4k}). \tag{13}$$

Thus the dynamics can be thought as a particle moves in the potential field $V(R)$ with mass $M = 2$ and total energy $E = -1$

$$\frac{1}{2}M(\frac{dR}{d\tau})^2 + V(R) = E. \tag{14}$$

Note that the potential energy depends on the initial velocity of the shell. The acceleration equation is

$$\frac{d^2R}{d\tau^2} = -\frac{m}{R^2} + 2kC(r)R^{4k-1}. \tag{15}$$

Define Schwarzschild-like coordinates $(T, R)$ as [20][24]

$$ds^2 = B(R,T)dT^2 - A(R,T)dR^2 - R^2d\Omega^2, \tag{16}$$

where $T$ is the clock time at infinity. According to Birkhoff's theorem, in the region $r > b$, the metric is

$$ds^2 = (1-\frac{2m(b)}{R})dT^2 - (1-\frac{2m(b)}{R})^{-1}dR^2 - R^2d\Omega^2. \tag{17}$$

Similarly, in the region $r < a$, under the requirement that the metric should be continuous, the metric takes the following form

$$ds^2 = f(T)(1-\frac{2m(a)}{R})dT^2 - (1-\frac{2m(a)}{R})^{-1}dR^2 - R^2d\Omega^2, \tag{18}$$

where $f(T)$ is a non-negative function.

Using the fact that $g^{TR} = 0$, we get

$$\frac{T'}{\partial T/\partial \tau} = R'\dot{R}(\frac{r}{R})^{4k}. \tag{19}$$

The characteristic equation of equation (16) is

$$\frac{d\tau}{dr} = -R'\dot{R}(\frac{r}{R})^{4k}. \tag{20}$$

The metric on the outer boundary can be obtained by inserting $r = b$ in equation (17),

$$(d\tau^2)_{r=b} = (1-\frac{2m(b)}{R(\tau,r=b)})dT^2 - (1-\frac{2m(b)}{R(\tau,r=b)})^{-1}(\frac{\partial R}{\partial \tau})^2(\tau,r=b)d\tau^2. \tag{21}$$

From this condition we obtain a boundary condition for $T(\tau,r)$, which reads

$$T(\tau,r=b) = \int \sqrt{(1-\frac{2m(b)}{R(\tau,r=b)})^{-1}(1+(1-\frac{2m(b)}{R(\tau,r=b)})^{-1}(\frac{\partial R}{\partial \tau})^2(\tau,r=b))}d\tau. \tag{22}$$

In view of (11), we have

$$T(\tau, r=b) = \int (1 - \frac{2m(b)}{R(\tau, r=b)})^{-1} \sqrt{C(r=b)} R^{2k}(\tau, r=b) d\tau. \qquad (23)$$

Integrate equation (20) with the boundary condition in equation (23), we can obtain $T = T(R, r)$.

### 3. Special solutions
#### 3.1 Analytical solutions in comoving coordinates
Assume that at the time $\tau = 0$ we have the scaling condition

$$R(0, r) = r. \qquad (24)$$

Choose the unknown function in equation (11) to be

$$C(r) = \frac{1}{r^{4k}}. \qquad (25)$$

This is equivalent to choose the initial velocity as

$$\frac{dR}{d\tau}(r, \tau=0) = \pm\sqrt{\frac{2m(r)}{r}}. \qquad (26)$$

For collapsing shells, we should choose minus sign in (26). Under such initial conditions, our special solutions coincide with Liu and Zhang's [2] for dust shells as shown below.
Taking into account of those conditions, equation (11) becomes

$$(\frac{dR}{d\tau})^2 = \frac{2m}{R} - 1 + \frac{1}{r^{4k}} R^{4k}. \qquad (27)$$

This equation can also be written as the form of (14), the potential function is

$$V(R) = -(\frac{2m}{R} + \frac{1}{r^{4k}} R^{4k}). \qquad (28)$$

Consider a shell collapsing towards a black hole with the following mass distribution

$$m(r) = m_B + c(r^3 - a^3), \tag{29}$$

where $m_B$ is the mass of the pre-existing black hole; the above results are reduced to that in Ref. [11] when $m_B = 0$ and $a = 0$

Integrate equation (27), then we obtain the evolution equation of area radius $R$ for a fixed particle ($r = $ const.)

$$\int_r^R \frac{dR}{\sqrt{\frac{2m}{R} - 1 + \left(\frac{R}{r}\right)^{4k}}} = \pm \int_0^\tau d\tau, \tag{30}$$

here "$+$" and "$-$" correspond to the outgoing and ingoing motion of fluid particles, respectively.

When $k = 0$, the solution is

$$\pm \tau = \frac{2}{3\sqrt{2m(r)}} (R^{3/2} - r^{3/2}). \tag{31}$$

When $k = 1/4$, there is an exact solution

$$\pm \tau = \sqrt{2r(r + \sqrt{r(r-8m)})} \left\{ F\left(\text{Arcsin}\left(\frac{1}{2}\sqrt{\frac{r + \sqrt{r(r-8m)}}{m}}\right), \sqrt{\frac{r - \sqrt{r(r-8m)}}{r + \sqrt{r(r-8m)}}}\right) - \right.$$

$$E\left(\text{Arcsin}\left(\frac{1}{2}\sqrt{\frac{r + \sqrt{r(r-8m)}}{m}}\right), \sqrt{\frac{r - \sqrt{r(r-8m)}}{r + \sqrt{r(r-8m)}}}\right) -$$

$$F\left(\text{Arcsin}\left(\frac{1}{2}\sqrt{\frac{R}{r}}\sqrt{\frac{r + \sqrt{r(r-8m)}}{m}}\right), \sqrt{\frac{r - \sqrt{r(r-8m)}}{r + \sqrt{r(r-8m)}}}\right) +$$

$$\left. E\left(\text{Arcsin}\left(\frac{1}{2}\sqrt{\frac{R}{r}}\sqrt{\frac{r + \sqrt{r(r-8m)}}{m}}\right), \sqrt{\frac{r - \sqrt{r(r-8m)}}{r + \sqrt{r(r-8m)}}}\right) \right\}, \tag{32}$$

where $F(\theta, k) = \int_0^\theta \frac{d\theta}{\sqrt{1-k^2 \sin^2 \theta}}$ is the incomplete elliptic integral of the first kind, $E(\theta, k) = \int_0^\theta \sqrt{1-k^2 \sin^2 \theta} \, d\theta$ is the incomplete elliptic integral of the second kind.

When $k = -1/4$, there is also an exact solution

$$\pm\tau = (2m+r)\operatorname{Arcsin}(\frac{\sqrt{2mR}-\sqrt{(2m+r-R)r}}{2m+r}) + \sqrt{2mr} - \sqrt{(2m+r-R)R}. \quad (33)$$

3.2 Qualitative features of the special solutions in comoving coordinates

So far we have obtained the analytical solutions for shells with three different values of $k$. In this section, we make some qualitative descriptions of the motion of shells in comoving coordinates for three typical scenarios, which represent the three possible endings of collapsing shells.

The first possible scenario is that a shell collapses eventually onto the pre-existing black hole within finite proper (comoving) time, as illustrated in Fig. 1 for the potential function (in equation (28)) of a shell with positive, negative or vanishing tangential pressure. Note that collapsing eventually onto the pre-existing black hole is the only possible ending for a shell with vanishing or negative tangential pressure, since the potential function decreases monotonically as the shell collapses, i.e., $R$ decreases. In order for a shell with positive tangential pressure to end this way, the maximal value of the potential function should be less than the total energy $E = -1$, as shown in Fig. 1, or the initial point lies on the left side of the maximal point.

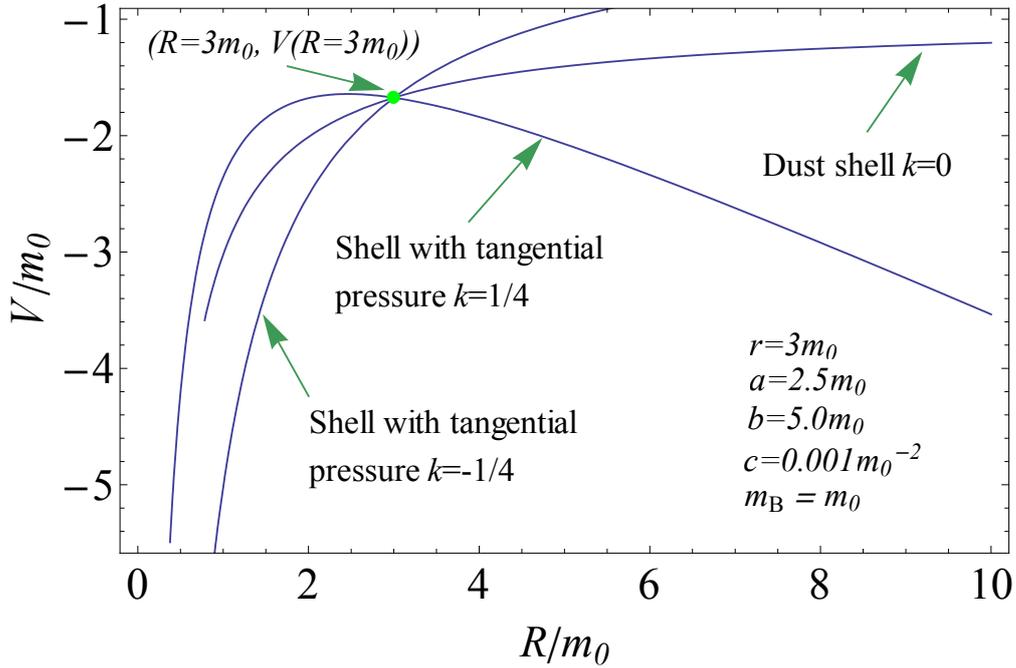

**Fig. 1.** The potential functions of shells collapsing eventually onto a black hole with three different values of $k$. $(R = 3m_0, V(R = 3m_0))$ is the common initial position of the shells.

The second possible scenario is that an initially collapsing shell escapes eventually from the gravitational field of the pre-existing black hole to infinity. This is possible only for shells with positive tangential pressure. In this case, the maximal value of the potential function must be larger than $E = -1$ and the initial position should be on the right side of the maximal point of the potential function, as shown in Fig. 2, where any motion on the potential function above $E = -1$ are not allowed. The left part of the maximal point of the potential function cannot be reached by the shell in this case. When the shell collapses initially from any point on the right side of the maximal point of the potential function, the shell will be reflected at point A and then expand to infinity eventually, due to the repulsive force from positive pressure.

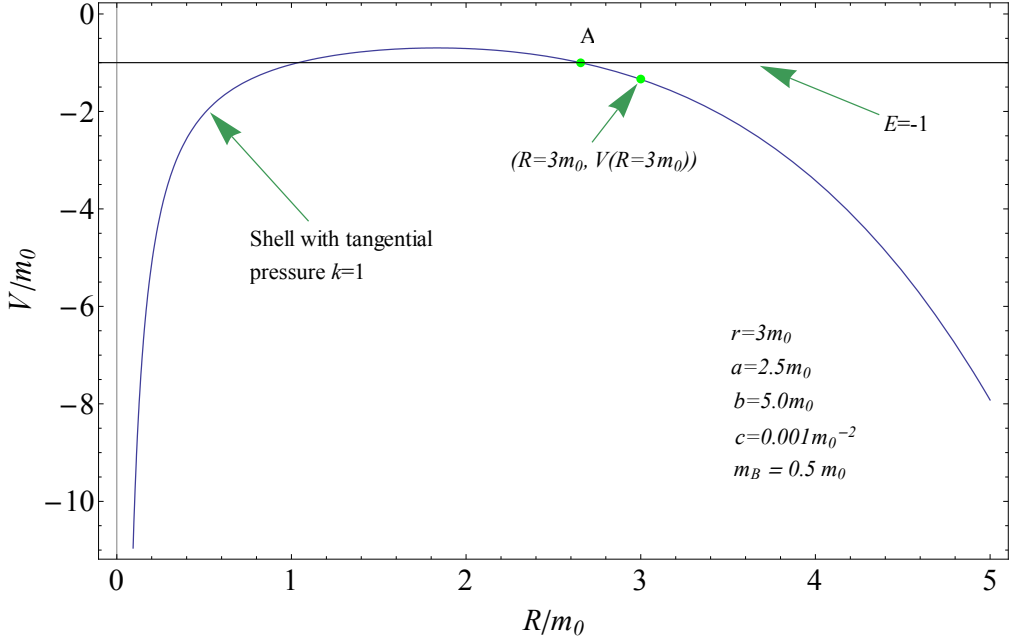

**Fig. 2.** The potential function of a shell around a black hole with $k = 1$: an initially collapsing shell from the right side will be reflected at point A and escapes to infinity eventually. $(R = 3m_0, V(R = 3m_0))$ is the initial position.

The last possible scenario is between the above two, i.e., the shell with positive tangential pressure neither falls onto the black hole nor escapes from the black hole, that is, the shell will be static at a particular point, which is a static solution to equation (11). For such a solution, both the velocity and acceleration must vanish, namely

$$\frac{dR}{d\tau} = 0, \tag{34}$$

and

$$\frac{dV}{dR} = 0. \tag{35}$$

According to equations (11) and (15), these two conditions can also be written as

$$R^{4k} = \frac{1}{(4k+1)C(r)}, \tag{36}$$

and

$$m(r) = 2kC(r)R^{4k+1}. \tag{37}$$

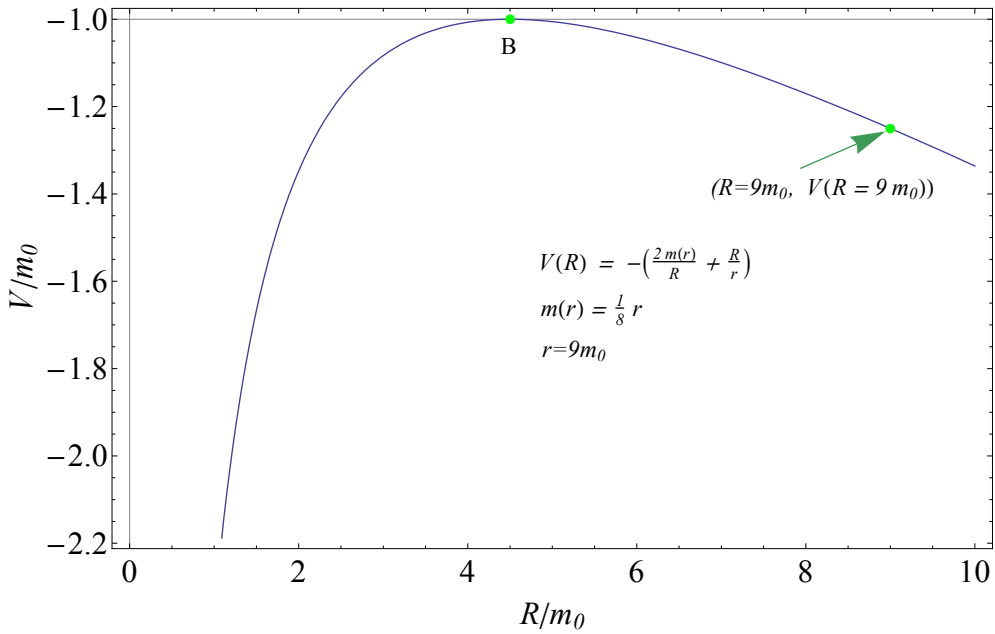

**Fig. 3.** The static solution for a shell around a black hole. $(R = 9m_0, V(R = 9m_0))$ is the initial point, and B is the static, but unstable point.

These static solutions are all unstable, because such a solution has a maximal potential energy at the static point, that is

$$\frac{d^2V}{dR^2} = -(\frac{4m}{R^3} + 4k(4k-1)C(r)R^{4k-2}) = -4k(4k+1)C(r)R^{4k-2} < 0, \tag{38}$$

for $k > 0$. Such a scenario is illustrated in Fig. 3 by choosing a potential function with maximal value $E = -1$; the shell starting from the initial point will eventually stop at the static point B. However, the shell will leave this static position under any small perturbation due to the unstable nature of the static solution; the shell will either fall onto the black hole or expand to infinity eventually.

## 3.3. Numerical results for collapsing shells

The motion of a shell around a black hole in comoving coordinates is described by (31), (32) and (33). In this section, we will only consider the scenario that a shell will collapse onto the pre-existing black hole eventually in the comoving coordinates, i.e., the first scenario discussed in section 3.2. Our goal is to find out exactly how the shell evolves in Schwarzschild coordinates, i.e., clocked by a distant stationary observer, by integrating the characteristic equation (20) with boundary condition (23).

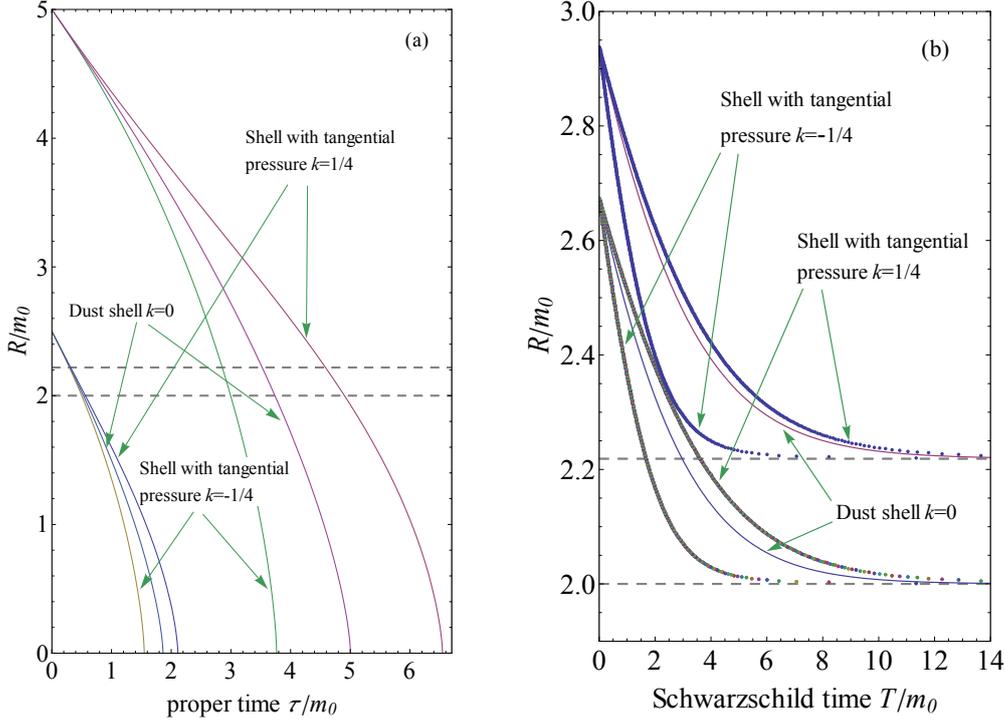

**Fig. 4.** The left panel (a) shows the evolution curves for a dust shell and a shell with tangential pressure in the comoving coordinates. The right panel (b) shows the evolution curves for a dust shell and a shell with tangential pressure in the Schwarzschild coordinates. The upper dashed lines indicate the final event horizons, and the lower dashed lines indicate the initial event horizons. Here we choose $m_B = m_0$, $a = 2.5m_0$, $b = 5.0m_0$ and $c = 0.001m_0^{-2}$.

In Fig. 4, we show the evolution curves of collapsing shells with respect to proper time and Schwarzschild time, i.e., clocked by the comoving observer and the distant stationary observer, respectively. The parameters in this figure are thus the same with the ones in Fig. 1 and are chosen in such a way that the shell can falls onto the black hole in comoving coordinates and shell-crossing singularities will not appear.

As shown in the left panel of the figure, compared with the case of a dust shell, the shell with positive tangential pressure collapses slower due to the resistance from pressure, and the shell with negative tangential pressure collapses faster due to the attraction from negative pressure. According to the right panel of this figure, we can see that every part of the shell can cross the event horizon with finite Schwarzschild time $T$ except the outermost boundary, which approaches the event horizon with an arbitrarily small distance. It can be seen that the qualitative features of the evolution curves are similar for all three cases.

**4. Summary**

With the analysis presented above, we conclude that the gravitational collapse of shells with positive tangential pressure are different from shells with vanishing or negative pressure. For the latter two cases, a shell must collapse onto the pre-existing black hole eventually. However, for the former case, a shell may collapse onto or away from the black hole eventually, depending on the initial conditions of the shell and the parameters of the system. For all collapsing solutions in the comoving coordinates, even with positive or negative tangential pressure, the qualitative features of the motion of the shell are similar to a dust shell in both coordinates. Therefore, the conclusions in Ref. [2] are still valid in the presence of tangential pressure for shells with collapsing solutions, as follows. A shell can indeed cross the event horizon of the black hole from the point of view of a distant stationary observer if the shell does so in the comoving coordinates, even when taking tangential pressure into account. There will be only infinitesimal amount of matter remaining outside the event horizon eventually, which apparently contradicts the picture obtained when the mass of in-falling matter is not considered, i.e., the test particle scenario commonly used to describe mattering falling onto a black hole. It is easy to see that, in a real astrophysical setting, the whole shell will cross the event horizon completely, since we can mimic the outer part of the shell as all matter between the observer and the observed in-falling matter shell [2]. Such a solution is different from the so-called frozen star [26], in which case all in-falling matter is supposed to accumulate outside the event horizon in the reference system of a distant stationary observer. Therefore, in real astrophysical settings, contrary to the well-known phenomenon of frozen stars [26], black holes can indeed be formed and all in-falling matter (even with tangential pressure) can cross the event horizon of the pre-existing black hole within finite time, according to the clock of a distant stationary observer. In this sense we can observe the matter fall into a black hole and the frozen star paradox is solved. This implies that only gravitational wave radiation can be produced during the merging process of two black holes formed by matter collapses, which is very different from the case of two merging frozen stars formed by matter collapses, in which case electromagnetic waves may also be generated besides gravitational waves due to the accumulated matter outside their event horizons.

It is also worth noting that although the in-falling matter can indeed reach the singularity within finite time of the comoving observer, the in-falling matter cannot approach the singularity at the center if clocked by a distant stationary observer; this conclusion is the same as that in Ref. [2], because the time of the comoving observer cannot be mapped to that of the distant stationary observer after the event horizon crossing. Since the universe has a finite age, the clocks of all observers outside black holes must be finite and can be synchronized in principle, it thus can be concluded that no matter can reach the singularity point at the center of any astrophysical black hole formed through matter collapse after the birth of the universe, within the finite clock time of any observer outside any black hole. For example, if an astronaut leaves the earth to travel to a black hole and eventually falls into the black hole, the astronaut will end up at the central singularity point within his finite time, as shown in Figure 4 (a); however, the final crash never happens to us on the earth and he/she will approach indefinitely to a finite location between the event horizon and the central singularity point, as shown in Figure 4 (b). Since the only known mechanism of forming macroscopic astrophysical black holes in the universe is through matter collapse, we can further conclude that there is nothing at the central singularity point within any such black hole, for all observers outside any such black hole. And finally, since the only black holes known so far are macroscopic astrophysical black holes, it is inevitable that astrophysical singularity does not exist in this universe, as far as any observer outside any black hole is concerned. Black hole singularity is thus simply a property of a black hole for a comoving observer who has already vanished into a black hole, but not an astrophysical reality for observers outside black holes [27, 28].

Though our solutions are for idealized cases and the radial pressure is also omitted in our setting, we consider that our conclusions are generally true for all practical situations of matter falling towards black holes. The fundamental reason is that the event horizon expands when the mass of an in-falling shell is taken into account for all collapsing solutions. On the contrary, the mass of in-falling matter has been ignored in many previous studies that considered the in-falling mass as a test particle, which cannot influence the metric of the whole gravitating system [2].

**Acknowledgments**

Anonymous referees are thanked for their comments and suggestions. Financial supports are given in the footnote on the first page. We thank Dr. Yuan Liu for many useful discussions and suggestions, as well as reading the draft of this manuscript.